\documentclass[%
 reprint,
 amsmath,amssymb,
 aps,
]{revtex4-2}

\usepackage{graphicx}
\usepackage{dcolumn}
\usepackage{bm}
\usepackage{hyperref}
\hypersetup{
  colorlinks   = true, 
  urlcolor     = blue, 
  linkcolor    = red, 
  citecolor   = red 
}

\begin{document}

\preprint{APS/123-QED}

\title{New scattering zones in quantum speckle propagation}

\author{Shaurya Aarav} \thanks{Current affiliation: Sorbonne Universit\'e, CNRS, Institut des NanoSciences de Paris, INSP, F-75005 Paris, France} \author{S. A. Wadood}
\author{Jason W Fleischer}%
 \email{jasonf@princeton.edu}
\affiliation{Department of Electrical and Computer Engineering, Princeton University, NJ 08544, USA}%

\date{\today}

\begin{abstract}
Quantum speckles exhibit significantly richer behavior than their classical counterparts due to their higher dimensionality. A simple example is the far-field speckle pattern in 1D light scattering: classical light forms 1D speckles defined by the numerical aperture, whereas biphoton scattering depends in addition on the photon correlation length, forming 2D elliptical speckles. To date, the behavior of quantum speckles for shorter propagation distances has not been considered. We remedy this here by considering the paraxial evolution of two-photon entanglement at arbitrary propagation distances from an isotropic scatterer. We show, theoretically, numerically, and experimentally, that the two length scales of the biphoton introduce a new Fresnel regime between the conventional near and far fields. Further, we show that the quantum near field is characterized by speckles with a square shape that remain constant during propagation. In contrast, the intermediate regime can be engineered to have a constant speckle size along the sum coordinate but a linearly expanding speckle size along the difference coordinate, with a speckle shape that transitions from square to elliptical. The results merge quantum coherence with scattering statistics and suggest new regimes of operation for correlation-based quantum sensing and imaging. 
\end{abstract}

\maketitle


\section{\label{sec:level1}Introduction}
\noindent Speckle is the result of random interference, such as that caused by scattering media, and depends sensitively on both phase and statistics. While its use in vibrometry has been recognized for decades \cite{leendertz1970interferometric}, its presence in other fields has been considered detrimental. Even though the speckle intensity is random, with an increased understanding of the underlying correlations and an exponential increase in computing power, speckle has become appreciated as a resource for imaging tasks and information processing. Examples include enhanced focusing \cite{Vellekoop:07}, lensless imaging \cite{bertolotti2012non,katz2014non,antipa2018diffusercam}, 3D imaging, \cite{aarav2023using,aarav2024depth}, and reservoir computing \cite{rafayelyan2020large}.

Speckles can be correlated in different degrees of freedom, and their usefulness often depends on the particular problem. For example, local shift invariance is useful for memory-effect based imaging of 2D objects \cite{freund1988memory}, intensity statistics has utility in optical manipulation \cite{Sun_PhysRevLett.108.263902,bender2018customizing}, and wavelength correlations allow for spectrum analysis \cite{metzger2017harnessing}. Here, we focus on the spatial correlations of speckles to highlight the differences between classical and quantum (biphoton) speckle propagation. We consider the simplest case: light of a single wavelength $\lambda$, propagation with one transverse dimension, and transmission through a thin, homogeneous scattering medium with correlation length $\sigma_0$ (the reflection case is similar).

In analogy with free-space propagation, the propagation of classical speckles away from a scattering medium is conventionally defined by a near field and a far field. In the latter, light scattered from different regions of the medium can overlap, resulting in a speckle size that grows linearly with distance from the scatterer (a direct consequence of the van Cittert-Zernike theorem \cite{goodman2007speckle}). In the former (sometimes called the deep Fresnel regime), different regions propagate individually, and the speckle size stays constant. The crossover distance is determined by the correlation length $\sigma_0$ and the numerical aperture \cite{giglio2000space,gatti2008three,magatti2009three,cerbino2007correlations}. 

To date, studies of biphoton speckle propagation have been restricted to the far field, where the main observed difference from the classical case is the shape of the speckle \cite{peeters2010observation,soro2021quantum}. This is because entangled photon pairs have an extra degree of freedom, the relative separation $\sigma_-$ of their components, other than the overall beam size $\sigma_+$. The ratio of the two sizes causes the speckles to appear elliptical while still maintaining the linear scaling with propagation, like in the classical case.

Here, we show that the biphoton speckle exhibits even richer behavior when the propagation distance is shorter. Most significantly, quantum correlation introduces a new regime of speckle propagation, intermediate between the near and far fields. In this region, the sum coordinate correlation of the photon pairs can propagate without change, while the difference coordinate grows linearly. Moreover, the shape of the speckle is not, in general, elliptical. When both coordinates propagate unchanged, i.e. in the near field, the speckle shape is square. It is only in the far field that speckles are truly elliptical. We support our theoretical results using simulations and experiments and discuss the implications of these results for higher photon number states.

\begin{figure*}[t]
\centering
\includegraphics[keepaspectratio=true, scale = .74,trim = {2.1cm 11.cm 0cm 8cm},clip]{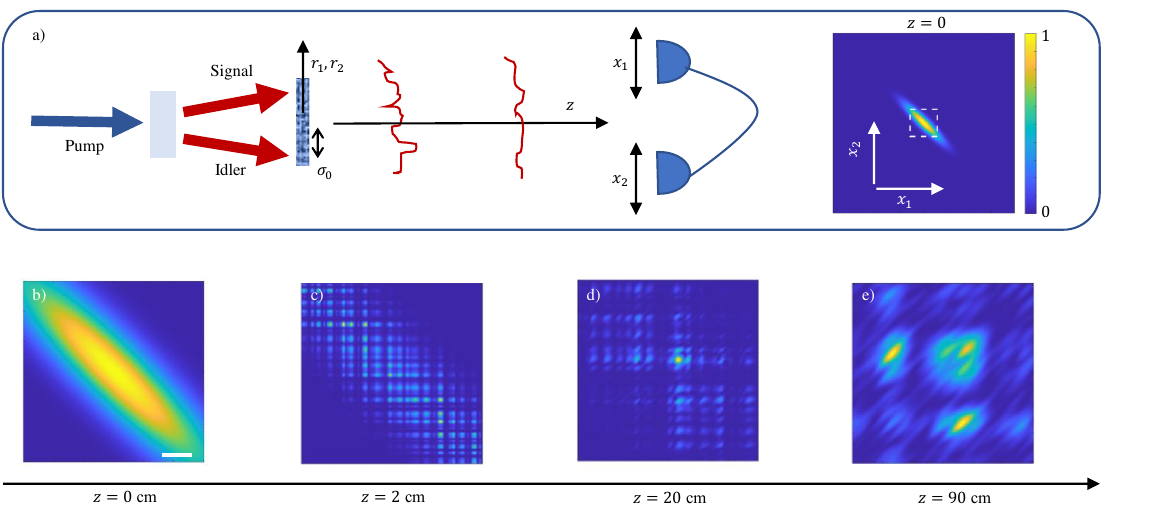}
\caption{\label{fig_sim_speckles} Simulation of biphoton speckles with propagation. (a) Simulation scheme. Biphotons scatter from an isotropic medium with a correlation length $\sigma_0$, propagate a distance $z$, and are measured in coincidence. Right: biphoton intensity at the scatterer plane $z=0$. The dashed square indicates the cropped region of interest for the bottom row. (b-e) Evolution of the biphoton intensity with $z$. (b) Initial intensity. (c) At small propagation distances, the speckles are squarish in shape and appear in a grid-like lattice oriented along the anti-diagonal. (d,e) As $z$ increases, the large-scale grid structure becomes more discrete and the speckle shape becomes more elliptical, with individual orientations along the diagonal.}
\end{figure*}

\section{Theory}

\noindent A schematic of our setup is shown in Fig. \ref{fig_sim_speckles}(a). Biphotons are generated by spontaneous parametric down-conversion (SPDC) \cite{karan2020phase}, propagate through a thin scattering medium, and are measured in coincidence. Let $\psi_{in}(r_1,r_2)$ be the biphoton field just before the scatterer, where $(r_1,r_2)$ are the positions of photons $1$ (signal) and $2$ (idler) on the scatterer plane. Under the Gaussian approximation, $\psi_{in}$ can be written \cite{fedorov2009gaussian,schneeloch2016introduction}

\begin{align}
\label{df_input}
    \psi_{in} = e^{-\frac{(r_1-r_2)^2}{\sigma_-^2}}e^{-\frac{(r_1+r_2)^2}{\sigma_+^2}}
\end{align}

\noindent where $\sigma_-, \sigma_+$ depend on the pump profile and crystal thickness, repectively \cite{karan2020phase}. The specific Gaussian form is not essential for our subsequent analysis; we show it above to emphasize the existence of two different length scales in the input state. Typically, $\sigma_- < \sigma_+$ in the near field and $\sigma_- > \sigma_+$ in the far field. Without loss of generality, we operate in the near field of the crystal. Physically, $\sigma_+$ determines the overall spatial extent of the beam when measuring single photons, while $\sigma_-$ determines the extent of the signal photon when conditioned on the coincidence measurement of the idler photon. In the far field of the crystal, our results remain the same except for the swapping of the two parameters $\sigma_-\leftrightarrow \sigma_+$. For simplicity, we consider 1D photons, so that the biphoton field is 2D.

Due to the thin scatterer, each photon acquires a phase $\phi(r)$ such that the biphoton field after the scatterer $\psi_0 = \psi_{in}(r_1,r_2) e^{i\phi(r_1)} e^{i\phi(r_2)}$. Upon propagating a distance $z$, $\psi_0 \rightarrow \psi_z$, which we model using the Fresnel approximation. We study the spatial structure of speckles using the biphoton correlation function $\Gamma_z = \langle\psi_{z}(r_1,r_2)\psi_{z}^*(r_1',r_2') \rangle$, where $\langle \cdot \rangle$ represents an ensemble average over many scatterer realizations and $^*$ denotes complex conjugate. See the supplement for more details.

\begin{figure*}[t]
\centering
\includegraphics[keepaspectratio=true, scale = .76,trim = {-1.5cm 15.cm 0cm 8.cm},clip]{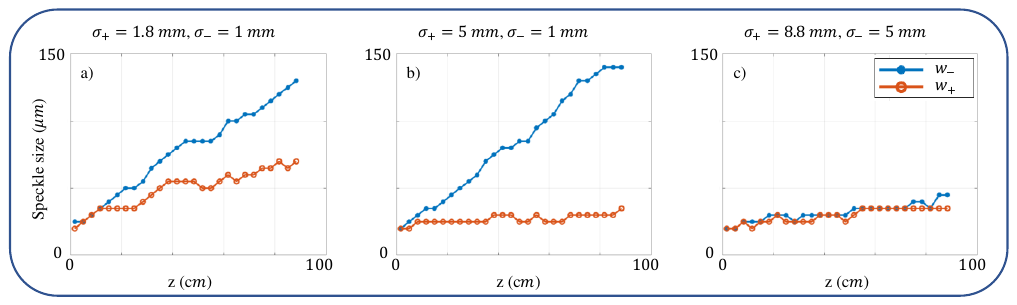}
\caption{\label{fig_sp_size} Simulations of biphoton speckle size as a function of propagation distance $z$. The wavelength and scattering correlation length are fixed at $\lambda = 810~\textrm{nm}$ and $\sigma_0 = 44~ \mu\textrm{m} $, while the biphoton input parameters $\lbrace \sigma_-, \sigma_+ \rbrace$ are chosen such that the speckles are
(a) in the far-field zone for the full range of $z$; (b) in the intermediate zone; and (c) in the near-field zone.}
\end{figure*}

To make the calculation of $\Gamma_z$ tractable, we follow \cite{peeters2010observation} and make two assumptions. First, we assume that the scattered biphoton field exhibits Gaussian statistics, so that the correlation of the field and intensity are directly related. Second, we assume that the scatterer is strong, such that the correlation function at $z\rightarrow0$ is 
\begin{align}
\label{df_input_corr}
    \Gamma_0 \approx R_{0}(\bar{r}_+,\bar{r}_-)\mu(\delta r_1,\delta r_2)
\end{align}

\noindent where $R_{0}(\bar{r}_+,\bar{r}_-)$ is the average biphoton intensity determining the coincidence for $\psi_{in}$ and $\mu(\delta r_1,\delta r_2)$ is the correlation function of the scatterer. The width of $\mu$ is the scatterer correlation length $\sigma_0$, assumed the same along both coordinates. Details of the coordinate transforms are listed in Table \ref{df_coords}.

Apart from the wavelength $\lambda$, there are 3 other length parameters: $\sigma_-, \sigma_+,$ and $\sigma_0$. Their combinations divide the axial direction into distinct regions, observed in the $z$-evolution of the correlation function $\Gamma_z$. We assume a separation of length scales, such that the scatterer correlation length $\sigma_0 <<\sigma_-, \sigma_+$. The axial regions of propagation then become:

\medskip

\noindent 1) \textit{Far field: $z>\sigma_0\sigma_+/\lambda$}
\begin{align}
\label{df_output_corr_FF}
    \Gamma_z = C &\mathfrak{F}[R_{0}(\bar{r}_+,\bar{r}_-)]\left(\frac{1}{z\lambda}(\delta x_+,\delta x_-)\right) \\ \nonumber
    &\times \mathfrak{F}[\mu(\delta r_1,\delta r_2)]\left(\frac{1}{z\lambda}(\bar{x}_+,\bar{x}_-)\right),
\end{align}

\noindent where $C$ is a phase factor and $\mathfrak{F}$ denotes a Fourier transform. This result recovers the far-field speckle correlations obtained in \cite{peeters2010observation},  with elliptical speckles with widths $w_+ = z\lambda/\sigma_+$ and $w_- = z\lambda/\sigma_-$ along their respective axes. The field of view is determined by the scatterer properties and is isotropic, giving the $\pm$ widths $l_+ = l_- = z\lambda/\sigma_0$.

Note that the defining distance of the far-field regime is similar to that of single-photon speckles, with the beam width $\sigma_+$ replaced by the aperture size \cite{cerbino2007correlations,gatti2008three,magatti2009three}. This width also determines the global properties of the beam, e.g. its field of view. The local properties, e.g. the speckle shape, are drastically different due to the biphotons being twice the dimension of single-photon speckles; elliptical speckles cannot exist in 1D.

\begin{table}[t]
\centering
\caption{Coordinate transforms.}
\vspace{.2cm}
\begin{tabular}{ |c c c|}
 $r_1 = \bar{r}_1+\delta r_1$ & \& & $r_1' = \bar{r}_1-\delta r_1$  \\ 
 $r_2 = \bar{r}_2+\delta r_2$ &\& & $r_2' = \bar{r}_2-\delta r_2$  \\ [1ex]
\hline
 $\bar{r}_+ = \bar{r}_1+\bar{r}_2$ &\& & $\bar{r}_- = \bar{r}_1-\bar{r}_2$  \\ 
 $\delta r_+ = \delta{r}_1+\delta{r}_2$ &\& & $\delta r_- = \delta{r}_1-\delta{r}_2$  \\  
\end{tabular}
\label{df_coords}
\end{table}

\medskip

\noindent 2) \textit{Near field: $z<\sigma_0\sigma_-/\lambda$}
\begin{align}
\label{df_output_corr_DF}
    \Gamma_z = R_{0}(\bar{x}_+,\bar{x}_-) \mu(\delta x_+,\delta x_-)
\end{align}

\noindent This result is similar to the near-field relation derived for single photons in \cite{cerbino2007correlations,gatti2008three,magatti2009three}. In this region, the biphoton speckle size depends only on the scatterer correlation function. Since $\mu$ has the same width along both photon coordinates $\delta r_1, \delta r_2$, the speckles appear as \textit{squares} of width $w_+ = w_- = \sigma_0$. The field of view is determined by the input state and is elliptical, with the $\pm$ widths given by $l_+ = \sigma_+$ and $ l_- = \sigma_-$, respectively.

\medskip

\noindent 3) \textit{Intermediate field: $\sigma_0\sigma_-/\lambda< z <\sigma_0\sigma_+/\lambda$}

\smallskip

\noindent This regime does not exist for single-photon speckles, as they lack the internal degrees of freedom necessary for scale separation during propagation.

\section{Simulations}

\noindent Here, we simulate the system shown in Fig. \ref{fig_sim_speckles}(a). In what follows, we keep a constant wavelength $\lambda = 810 ~\textrm{nm}$ and scatterer correlation length $\sigma_0 = 44~ \mu \textrm{m}$ but vary the biphoton parameters $\sigma_- $ and $\sigma_+$. In all cases, we observe the evolution of coincidence functions from the initial elliptical Gaussian intensity $|\psi_0|^2$ at $z=0$ to the far field nearly a meter later. Details of the simulation are given in the Supplement. 

Figures \ref{fig_sim_speckles}(b-d) show propagation results for $\sigma_- = 1~\textrm {mm}$ and $\sigma_+ = 5~\textrm{mm}$. As predicted, the near field $z^{NF} < \sigma_0\sigma_-/\lambda = 5~ \textrm{cm}$ is characterized by a grid-like pattern of square-like speckles (Figs \ref{fig_sim_speckles}(c)) that smoothly elongate in the far field $z^{FF} > \sigma_0\sigma_+/\lambda = 27~ \textrm{cm}$. In the intermediate regime, the grid structure becomes more distinct and the orientation of the internal elements transitions to anti-diagonal correlations.

To make our observations on speckle shape more concrete, we plot the speckle sizes for a given scatterer realization and  different input state parameters (Fig. \ref{fig_sp_size}). For Fig. (\ref{fig_sp_size} a), the widths of the input states are $\sigma_+ = 1.8~\textrm{mm}$ and $\sigma_- = 1~\textrm{ mm}$, giving cross-over distances for the near-field and far-field zones of  $z^{NF} \sim 5~ \textrm{cm}$ and $z^{FF} \sim 9~\textrm{cm}$, respectively. Thus, we expect (and observe) the speckle size to be elliptical above these $z$ values, with major axis length $w_+\propto 1/\sigma_-$ and minor axis length $w_-\propto 1/\sigma_+$.

\begin{figure*}[t]
\centering
\includegraphics[keepaspectratio=true, scale = .76,trim = {2cm 8.5cm 0cm 7.5cm},clip]{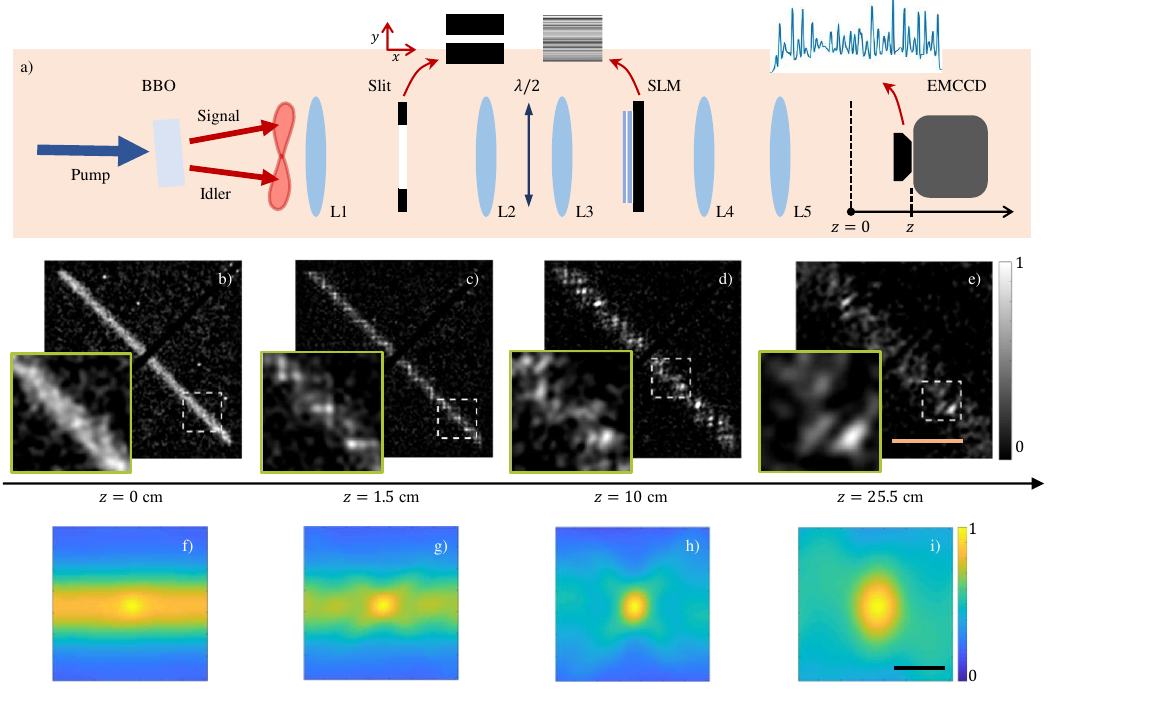}
\caption{\label{fig_expt_z} Experimental results. (a) Experimental setup. In the first half, biphotons created in a BBO crystal via Type-1 SPDC are shaped into a quasi-1D beam. In the second half, the beam is polarized by a half-waveplate, acquires a random “scattering” phase from an SLM, and is imaged into an EMCCD camera. Lens $L1$ performs an optical Fourier transform onto the slit, while lenses $\lbrace L2,L3 \rbrace$ and $\lbrace L4,L5 \rbrace$ are imaging relays. (b-e) Coincidence measurements at four propagation distances, depicting the evolution of biphoton speckles. Insets: magnified images of region within the dashed white box. Each speckle image is normalized by the maximum value in their inset. Scale bar: $2.7 ~\textrm{mm}$. (f-i) Speckle shape obtained by autocorrelation (cropped and rotated by $45^o$ for better visualization). Speckles start with a rectangular structure in (f), but transition to an elliptical feature in the far-field in (i). (g) and (h) show intermediate regimes where the speckle is neither rectangular nor completely elliptical. Scale bar: $0.37~\textrm{mm}$.}
\end{figure*}

In Fig. \ref{fig_sp_size}(b), we increase the width of the input state to $\sigma_+ = 5 ~\textrm{mm}$ while keeping the difference coordinate $\sigma_- = 1 ~\textrm{mm}$ the same. The near-field zone $z^{NF} \lesssim 5~\textrm{cm}$ thus remains the same, while the far-field zone increases to $z^{FF} \gtrsim 27~\textrm{cm}$. This corresponds to the middle region between the near field and the far field, with the speckle size along one dimension increasing linearly with $z$ and the speckle size along the perpendicular direction remaining flat. However, we do not see the expected transition to the far field zone at $z = 27~\textrm{cm}$. This is because we calculated the critical value $z^{FF}$ from a length-scale argument; in practice, the transition $z^{FF}$ will depend on the exact functional forms of the scatterer correlation and the input state, i.e. the exact results from the full Fresnel integral \cite{reichert2017quality}.

In Fig. \ref{fig_sp_size}(c), we demonstrate the speckle behavior in the near-field zone. To represent this region, we use larger values of both biphoton widths of the input state: $\sigma_+ = 8.8 ~\textrm{mm}$ and $\sigma_- = 5 ~\textrm{mm}$, giving critical distances $z^{NF} \sim  27 ~\textrm{cm}$ and $z^{FF} \sim 47 ~\textrm{cm}$. Since $\sigma_-$ now has the same width that $\sigma_+$ had in Fig. \ref{fig_sp_size}(b), we expect the same flat behavior with $z$. Indeed, Fig. \ref{fig_sp_size}(c) shows that the widths in both directions remain flat. This behavior corresponds to the square speckles in Fig. \ref{fig_sim_speckles}(c) and is characteristic of quantum light propagation in the near-field region.

\section{Experiments}
\noindent Our experimental setup is shown in Fig. \ref{fig_expt_z}(a). A $ 5~\textrm{mm} \times 5~\textrm{mm} \times 3~\textrm{mm}$ BBO (Beta Barium Borate) crystal is illuminated by a $405 ~\textrm{nm}$ laser pump. The crystal is rotated slightly to operate in the collinear regime and placed at the back focal plane of lens $L1$ with focal length $10 ~\textrm{cm}$. This lens images the beam onto a slit, which reshapes it to a ~1D profile, before acquiring a random phase imprinted by a spatial light modulator (SLM). We perform coincidence imaging of the biphotons using an Andor iXon Ultra EMCCD camera and use the scheme outlined in \cite{defienne2018general,reichert2018massively,defienne2022pixel,bhattacharjee2022propagation} by taking $1-2$ million frames at different exposure times. Different propgation distances are recorded by moving the camera axially.

Coincidence images and their autocorrelations are shown in Fig. \ref{fig_expt_z} (b-i). At $z=0$ (Fig. \ref{fig_expt_z} (b)), we see the anti-correlation diagonal  typical of the far field of an SPDC setup. The effect of the phase scattering is not yet visible, as this is the image plane of the SLM. Moving the EMCCD to $z = 1.5 ~\textrm{cm}$ (Fig. \ref{fig_expt_z} (c)), we observe square-shaped speckles appearing in a rectangular grid-like pattern, indicative of the quantum near-field region. The speckle shape can be better seen in the autocorrelation image (g). Moving the detector to $z= 10 ~\textrm{cm}$ (Fig. \ref{fig_expt_z} (d)), we see the speckles become somewhat elliptical, confirmed in the autocorrelation image (h). Lastly, we moved the detector to $z = 25.5 ~\textrm{cm}$ (Fig. \ref{fig_expt_z} (e)). The speckles now are markedly elliptical (again better visualized in the autocorrelation (i)), demonstrating the transition to the far-field region. 

\section{Discussion}

\noindent Because of the added degree of freedom, biphoton dynamics do not follow simply from quantization of classical near-field results. Coupled with propagation, the high-dimensional nature of biphotons enables entirely new speckle applications. In the near field, the use of biphotons instead of classical light can improve the capabilities of existing speckle measurement techniques, e.g. lensless imaging \cite{antipa2018diffusercam}, depth sensing \cite{aarav2023using}, and material characterization \cite{giglio2000space}. In the intermediate zone, the advantages of quantum imaging (such as higher resolution \cite{defienne2022pixel}, increased sensitivity \cite{reichert2017quality}, and improved noise properties \cite{brida2010experimental}), suggests the intriguing possibility to probe independently, and simultaneously, the near- and far-field speckle behavior of a scatterer. The near-field and intermediate zones also offer possibly novel connections between speckle statistics and entanglement \cite{pires2012statistical,beenakker2009two} and associated improvements to adaptive quantum optics \cite{defienne2018adaptive,lib2020real}.\par

There are significant implications of our work for higher-order photon correlations, which are essential for metrology \cite{dowling2008quantum}, imaging \cite{giovannettishapiro2009sub,defienne2024advances}, and computing \cite{aaronson2011computational}. While classical single-photon speckles cannot uniquely quantify biphoton speckle statistics, the latter do contain information about higher-order speckles. For example, the biphoton wave function uniquely determines all higher-order coincidence probabilities (and hence the speckle statistics) of multimode squeezed states generated via downconversion \cite{mollow1973photon,massar_lantz2023multiphoton}. Because multiphoton correlations can offer $N$-fold enhancements in phase sensitivity \cite{hiekkamaki2022observationofquantumgouyphase} and both transverse and axial resolution \cite{defienne2024advances,giovannettishapiro2009sub}, multiphoton speckles can offer more degrees of freedom and structure upon propagation. \par

\section{Conclusions}
\noindent In this work, we studied the propagation behavior of biphoton speckles. Unlike their classical counterparts, there exists a distinct intermediate propagation region between the near and far fields. The cross-over distances depend on the correlation length of the scatterer, the correlation lengths of the biphoton along its sum-difference coordinates, and the wavelength. Different ratios of these lengths correspond to differences in speckle shape, statistics, and dynamics. For higher-order photon number states, there are more internal degrees of freedom, with similar but more varied opportunities for separation of scales. Our work therefore informs the generation, propagation, and control over the scattering of high-dimensional entangled states. \par

\noindent \textit{Acknowledgments:} We thank Dr Abhinandan Bhattacharjee for the discussions regarding the measurement technique. This work was supported by AFOSR grants FA9550-19-1-0167 and FA9550-18-1-0219.


\bibliography{apssamp}

\clearpage
\onecolumngrid
\section*{Supplementary Material}
This supplement gives more details on the theory, simulation, and experiment presented in the main text.

\section{Theory}
\noindent The evolution of biphotons with propagation can be written

\begin{align}
\label{df_prop_state}
    \psi_z = \int\int \psi_{0}(r_1,r_2) h_z(r_1,x_1) h_z(r_2,x_2) dr_1 dr_2
\end{align}

\noindent where the free-space transfer function $h_z(r_j,x_j) = e^{-ik_0 (r_j-x_j)^2/2z}$, $k_0 = 2\pi/\lambda$, and $r_{1,2},x_{1,2}$ denote the input- and output-plane coordinates, respectively. The corresponding biphoton correlation function $\equiv \Gamma_z = \langle\psi_{z}(x_1,x_2)\psi_{z}^*(x_1',x_2') \rangle$ is thus

\begin{align}
\label{df_output_corr}
    \Gamma_z  =  C\int &\Gamma_0 e^{-i\frac{k_0}{2z} [(r_1-x_1)^2- (r_2-x_2)^2]} \times e^{i\frac{k_0}{2z} [(r_1'-x_1')^2+(r_2'-x_2')^2]}
    dr_1dr_2dr_1'dr_2' \nonumber\\
\end{align}

\noindent where $C = e^{-i\frac{k_0}{z}(\delta x_+\bar{x}_++\delta x_-\bar{x}_-)}$ is a phase term (expressed in terms of the $\pm$ coordinates listed in Table I of the main text). Using the decomposition $\Gamma_0 \approx R_{0}(\bar{r}_+,\bar{r}_-)\mu(\delta r_1,\delta r_2)$ (Eq. (2) of the main text) and integrating along $\delta r_+, \delta r_-$ gives

\begin{align}
\label{df_output_corr_1in}
    \Gamma_z = C\int & R_{0}(\bar{r}_+,\bar{r}_-) \mathfrak{F}[\mu(\delta r_1,\delta r_2)]\left(\frac{1}{z\lambda}(\bar{x}_+-\bar{r}_+,\bar{x}_--\bar{r}_-)\right)  \times e^{-i\frac{k_0}{z} [-\delta x_+\bar{r}_+-\delta x_-\bar{r}_-]}d\bar{r}_+d\bar{r}_-
\end{align}

\noindent Below, we use this expression to calculate Fresnel propagation in the near and far fields.

\subsection{Far field}
\noindent We define the far field such that the propagation distance: $z>\sigma_0\sigma_+/\lambda$. This assumption greatly simplifies Eq (\ref{df_output_corr_1in}). It leads to $\mathfrak{F}[\mu(\delta r_1,\delta r_2)](\frac{1}{z\lambda}(\bar{x}_+-\bar{r}_+,\bar{x}_--\bar{r}_-))\approx \mathfrak{F}[\mu(\delta r_1,\delta r_2)](\frac{1}{z\lambda}(\bar{x}_+,\bar{x}_-))$. This is applicable because the width of $\mathfrak{F}[\mu(\delta r_1,\delta r_2)]$ along the two dimensions is $z\lambda/\sigma_0$, whereas the width of $R_{0}(\bar{r}_+,\bar{r}_-)$ is $(\sigma_+,\sigma_-)$ along the two dimensions. Since we assumed $\sigma_-<\sigma_+$, the far-field condition implies that $\mathfrak{F}[\mu]$ is much broader than $R_0$ as a function of $(\bar{r}_+,\bar{r}_-)$, hence justifying to omit $\mathfrak{F}[\mu]$'s dependence on $(\bar{r}_+,\bar{r}_-)$. We can then integrate Eq (\ref{df_output_corr_1in}) to get:

\begin{align}
\label{df_output_corr_FF_s}
    \Gamma_z = C &\mathfrak{F}[R_{0}(\bar{r}_+,\bar{r}_-)]\left(\frac{1}{z\lambda}(\delta x_+,\delta x_-)\right) \times \mathfrak{F}[\mu(\delta r_1,\delta r_2)]\left(\frac{1}{z\lambda}(\bar{x}_+,\bar{x}_-)\right)
\end{align}

The above result recovers the far-field speckle correlations obtained in \cite{peeters2010observation}.

\subsection{Near field}
\noindent We define the near field region to be such that the propagation distance: $z<\sigma_0\sigma_-/\lambda$. Since the far-field assumption implied that $\mathfrak{F}[\mu]$ is much broader than $R_0[]$ as a function of $(\bar{r}_+,\bar{r}_-)$ in Eq (\ref{df_output_corr_1in}), the reverse is true here: $\mathfrak{F}[\mu]$ is much narrower along both dimensions here and samples $R_0$ effectively like a delta function in Eq (\ref{df_output_corr_1in}). It can be explicitly seen with the following transformation.

Let $ p = \frac{1}{z\lambda}(\bar{x}_+-\bar{r}_+), p'= \frac{1}{z\lambda}(\bar{x}_--\bar{r}_-) $. Eq (\ref{df_output_corr_1in}) becomes:

\begin{align}
\label{app_df_output_corr_1in}
    \Gamma_z =& C\int R_{0}(\bar{x}_+-z\lambda p,\bar{x}_--z\lambda p') \mathfrak{F}[\mu(\delta r_1,\delta r_2)](p,p') \nonumber\\
    &e^{i\frac{k_0}{z}(\delta x_+\bar{x}_++\delta x_-\bar{x}_-) - i2\pi\delta x_+p - i2\pi\delta x_-p'}dp dp' \nonumber\\
    \approx & R_{0}(\bar{x}_+,\bar{x}_-)\mu(\delta x_+,\delta x_-)
\end{align}

yielding the biphoton speckle correlation in the near field.

\section{Simulations}

\noindent The transverse size of the simulation is $2~\textrm{cm} \times 2~\textrm{cm}$, with each pixel covering an area of $5~\mu \textrm{m} \times 5 ~\mu \textrm{m}$. The wavelength of the biphoton is $810~\textrm{nm}$, as used in the experiment. The phase $\phi(x)$ of the 1D scatterer is generated by using random numbers in the range $(0,1)$ from a uniform distribution that are then stretched to cover the full phase range $[0,2\pi]$. Afterwards, we smoothen the profile with the `imgaussian' function in MATLAB and set  the scatterer correlation length to $\sigma = 44~\mu  \textrm{m}$ by fitting the autocorrelation of $e^{i\phi(x)}$ with a Gaussian. As shown in in Fig. (1) of main text, the widths of the input state are $\sigma_- = 1 \textrm{mm}$ and $\sigma_+ = 5 \textrm{mm}$. In our simulations, we use the angular spectrum method to propagate each photon \cite{goodman2005introduction}, which is more accurate than the paraxial approximation used in Eq. (\ref{df_prop_state}).

The blur filter size for all the results in Fig. 2 was fixed at $200~\mu \textrm{m}$. Before measuring the speckle size (determined by an autocorrelation), we removed the slowly varying envelope of the coincidence intensity by dividing the speckle image by its Gaussian blur. We then subtracted the mean, normalized with its maximum value, and put a binary threshold at $0.7$. Finally, for better visualization, we rotated the resulting image by $45$ degrees.

\section{Experiments}
\noindent The experimental setup is shown in Fig. 3(a) of the main text. Here, we continue its description, starting with the vertical slit used to control the overall spatial extent ($\sigma_+$) of the biphoton beam. The slit is imaged onto the plane of a Holoeye spatial light modulator (HED 6010 L NIR II) using two lenses: $L2$ of focal length $7.5~ \textrm{cm}$ and $L3$ of focal length $30~ \textrm{cm}$. The imaging relay lenses $L4,L5$ are both of focal lengths $20$ cm. To mimic an ideal thin scatterer, we applied a 1D random phase to the SLM, generated by the same algorithm used in the simulations. For the experiment, we performed a Gaussian blurring in MATLAB with filter size $=8$ and stretched the intensity values to the range ($0,255$) to match the grayscale bit depth of the SLM.

For coincidence imaging with the EMCCD, we used the scheme outlined in \cite{defienne2018general,reichert2018massively,defienne2022pixel,bhattacharjee2022propagation}. More specifically, we calculated the coincidences as

\begin{align}
    \Gamma_c(x_1,x_2) \propto  \sum_{k=1}^{N} I^{k}(x_1)I^{k}(x_2) -\sum_{k=1}^{N} I^{k}(x_1)I^{k+1}(x_2)
\end{align}

\noindent where $N$ is the number of frames and $I^{k}(x_1)$ represents the intensity obtained at pixel position $x_1$ in the $kth$ frame. To expedite the data acquisition, we limited data collection to 1D by performing horizontal binning in the camera itself. To remove noise, we blurred the measurements with a Gaussian filter and eliminated any negative values. We also removed any non-zero self-correlations by setting to zero the values on the right diagonal (and the neighboring $15$ pixels on its left and right sides).

Since the peak intensity decreases as the speckles propagate, we increased the exposure parameters for longer propagation distances. For $z = 0$ and $z=1.5~ \textrm{cm}$, we took $1.2$ million frames with an exposure time of $0.01~\textrm{s}$. For $z = 10~ \textrm{cm}$, the exposure time was $0.04$s with $2$ million frames. For $z = 25.5~ \textrm{cm}$, we took $2.2$ million frames with an exposure time of $0.04~\textrm{s}$. The filter sizes of the Gaussian blur for $z=0,1.5,10~ \textrm{cm}$ was $2.5$ pixels, while that for $z = 25.5~ \textrm{cm}$ was $5$ pixels.

\end{document}